# WAKEFIELD ACCELERATION OF SELF-INJECTED BUNCH IN A CONICAL PLASMA CHANNELS

*D. S. Bondar[1], W. Leemans[2], V. I. Maslov[1,2], and I. N. Onishchenko[1]*

[1]*National Science Center "Kharkiv Institute of Physics and Technology", Kharkiv, Ukraine*
[2]*Deutsches Elektronen-Synchrotron DESY, Hamburg, Germany*
*E-mail: bondar.ds@yahoo.com*

Laser wakefield acceleration is a widely studied method for accelerating charged particle bunches, with self-injection being a key feature. However, as the bunch accelerates beyond the driver velocity, it shifts out of the maximal accelerating wakefield phase. This work proposes using a tapering cone channel to address this issue. The gradual narrowing synchronizes the bunch with the wake phase by reducing the bubble size, keeping the bunch in the accelerating phase. The obtained dependencies of the bunch length, field $E_z$, and mean longitudinal momentum $p_z$ on the channel radius are useful for further researches.

PACS: 29.17.+w; 41.75.Lx

## INTRODUCTION

Laser-wakefield acceleration (the term laser-plasma acceleration - LPA is also often used) has grown from the first insight, who realized that a laser pulse could excite GV/m plasma waves capable of relativistic particle acceleration [1, 2, 3], into a mature method to obtain low size and high energy bunches with up to TV/m acceleration rates [4, 5, 6].

The use of plasma channels allows to increase the acceleration efficiency. The first experimental proof that centimeter-scale plasma channels can generate >1 GeV electrons came from the BELLA collaboration, where by using a guided 40-TW laser it was reached 1 GeV in 3 cm of capillary [7]. Multistage coupling then realized in [8] where shown that two independent LWFA cells, linked by a plasma lens, can save emittance and charge while doubling energy, establishing a modular roadmap for compact accelerators [9].

In [10-12] authors demonstrated a hybrid LWFA to PWFA configuration in which the self-injected bunch from the first stage drives a second, hundreds of microns-long plasmas, boosting gradients to ≈ 100 GV/m and more. Acceleration fields up to several hundred GV/m and up to TV/m were shown in [6, 13, 14, 15]. Self-injection control progressed through density shaping [16-19]. It is possible to reduce energy spread and emittance [20]. The advantage of conical plasma channels in increasing the energy of bunches has been shown previously [21], but the important effect of using the geometry of the conical channel to compress the wakefield bubble and ensure that the bunch is in the wakefield wave acceleration phase in a homogeneous plasma has not been studied enough.

In [22], the impact of structure geometry on electron acceleration was studied. Several configurations have been explored to improve beam transport and acceleration efficiency, including hollow plasma channels and plasma–dielectric waveguides [23], which provide focusing and eigenmode properties under bunch-driven excitation. In [24] experimentally demonstrated the generation of high-quality, quasi-monoenergetic relativistic electron bunches from an intense laser-plasma interaction by operating in a specific density regime just above the threshold for plasma wave-breaking. In [25] authors propose a method for direct laser acceleration in a plasma waveguide that allows for control over the laser field phase velocity for efficient energy gain by electrons. In [26] demonstrated that producing electron beams with needed properties is governed by a combination of two effects: beam loading, which terminates electron injection, and matching the accelerator length to the dephasing length to compress the bunch energy spread.

The study showed the advantage of a homogeneous conical plasma channel in a higher value of self-injected bunch longitudinal momentum compared to a homogeneous cylindrical channel. The study, the results of which are presented in this paper, were completed by numerical simulations using the code WarpX [27].

## STATEMENT OF THE PROBLEM

In Fig. 1 it was shown the configuration of conical plasma channels that were studied. The plasma inside the channel was uniform. Its density was $n_{e0}=1.74 \cdot 10^{19}$ cm$^{-3}$. In the figure, it is shown by a green fill. Colored lines show the radius of the conical channel in each of the cases under consideration. The following radii of the small base of the cone were considered: (1.26, 1.44, 1.62, 1.80, 1.98, 2.16, 2.34, 2.52, 2.70, 2.88, 3.06) $c/\omega_{pe}$. A cylinder with a radius of 4.00 $c/\omega_{pe}$ was also considered.

It is quite difficult to influence the properties of self-injected bunches. This can only be done indirectly - by changing the parameters of the laser and plasma. Increasing the longitudinal momentum of a self-injected bunch by holding the self-injected bunch at wakefield acceleration phase by reducing the size of the wakefield bubble in a narrowing conical channel is a method that would simplify the experiment.

## RESULTS OF SIMULATION

At Fig. 3 it was shown shapes of wake bubbles and the position of self-injected bunches. The sizes of the dots are different for better readability of the graph.

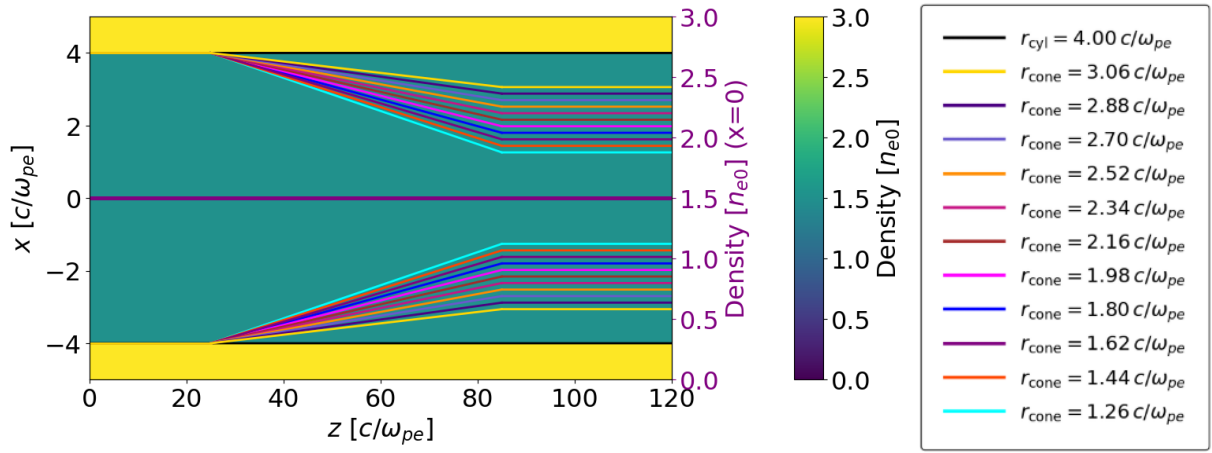

*Fig. 1. Plasma electron density profile $n_e(z, x)$. Longitudinal plasma density profile. Dynamic density change. The lines of different colors to mark the boundaries of cones of different radii; the cylinder is black line. The density outside the channels is $100\, n_{e0}$ (yellow).*

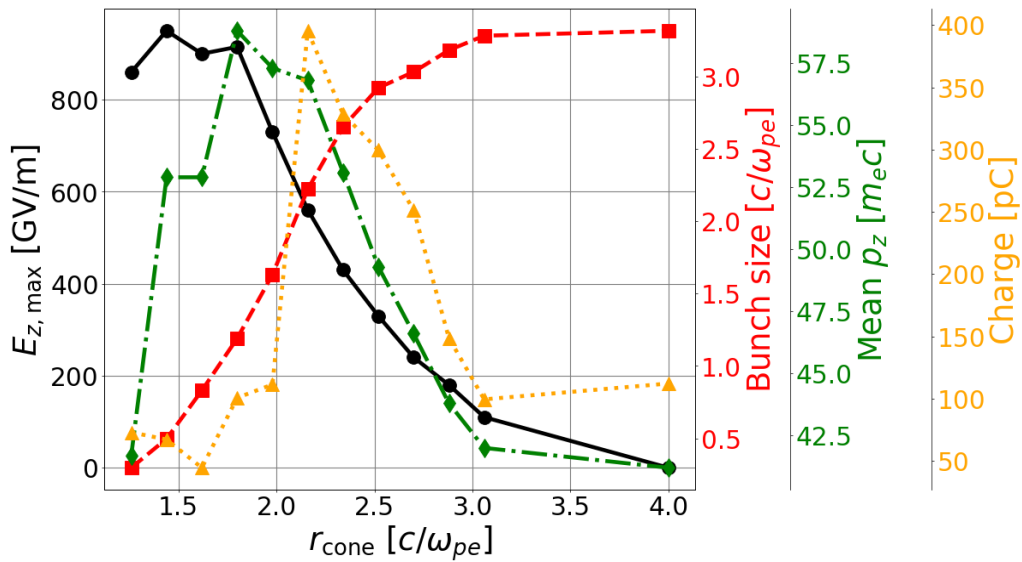

*Fig. 2. Dependence of the maximal acceleration field $E_{z,\,max}$ at the self-injected bunch region (black, circle marker); bunch size, length (red, square marker); mean $p_z$ (green, rhomb marker); mean $p_z$ (light orange, triangle marker) from the radius of the small base of the conical channel. The coordinate z corresponds to the exit of the conical channel. $r=4.00\, c/\omega_{pe}$ is cylinder. Time is 238 fs.*

Fig. 2 shows the main dependencies of the parameters of self-injected bunches and the accelerating field amplitude $E_z$ in the bunch region for different values of the radii of the small base of the cone. The parameters of the bunch and field were studied at moment t=238 fs, when the wakefield bubble is at the exit of the conical channel. Main parameters of the system: plasma density is normalized to $n_{e0}=1.74 \cdot 10^{19}$ cm$^{-3}$. The wavelength of the laser is $\lambda$=800 nm. Waist $w_0$=4.95 μm, duration $T_{full} \approx$ 30 fs. Units: $c/\omega_{pe}$=1.27 μm, $1/\omega_{pe}$=4.25 fs. $a_0=eE_0(m_e\omega c)^{-1}$=3. The spatial and temporal profiles of the laser are Gaussian. The channel walls are chosen so that they have a density $n_{e,\,walls}=100n_{e0}=1.74 \cdot 10^{21}$ cm$^{-3}$. The bunch current is normalized to the Alfven current $I_A=4\pi\varepsilon_0 m_e c^3 e^{-1} \approx$ 17 kA.

In Fig. 2 it can be seen that:

1) $E_{x,\,max}$ value decreases with increasing radius of the small base of the cone, i.e. as the shape of the conical channel approaches the cylindrical shape. This is due to the exit of the bunch from the acceleration phase of the wakefield - the main negative process that we are overcoming. However, other parameters should be taken into account in order to find the optimum;

2) the longitudinal size of the bunch (length) increases with radius increasing. This is due to the fact that in case of decreasing radius, the self-injected bunch falls to the back wall of the wakefield bubble and is partially and eventually completely absorbed by back wall;

3) the average value of the longitudinal momentum initially increases due to (due to the decrease in the size of the wake bubble) the holding of the self-injected bunch in the acceleration phase of the wakefield and at some point, falls due to the destruction of the bunch and its absorption by the back wall of the wakefield bubble. Fig. 3 shows the images of the wakefield bubble shapes in their own reference frames.

4) The behavior of the charge graph is related to the behavior of the density. In the cylindrical case, the bunch has a large longitudinal size, but its density is lower. Fig. 4, 5 show a self-injected bunch on graphs of density distribution (Fig. 4) and momentum $p_z$ (Fig. 5).

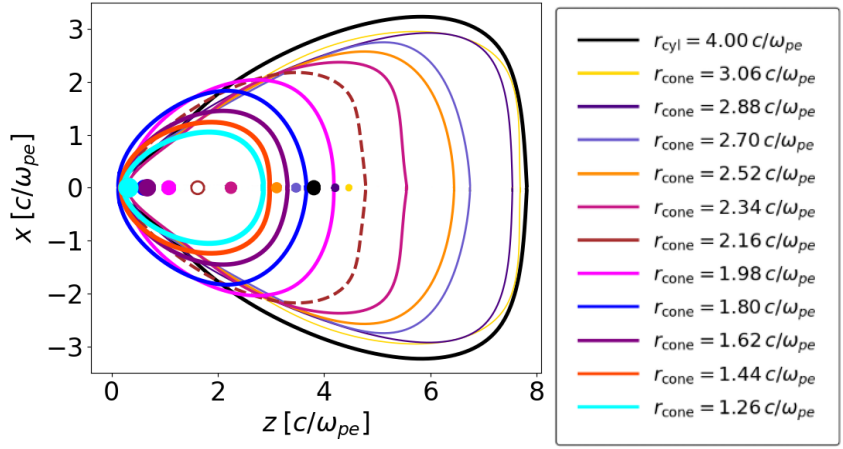

*Fig. 3. Dependences of the wake bubble size at the channel outlet on the cone (and cylinder) radius. The wake bubble is also shown for cylinder. The optimal case is shown in brown and dash ($r_{cone}$=2.16 $c/\omega_{pe}$, t=238 fs). Wake bubble reference system.*

The longitudinal momentum in fact illustrates the energy of the bunch due to the fact that the transverse momentum $p_x$ is close to 0.

The moments at which the bubble shape data were recorded were 238 fs after the start of the simulation to the moment the wake bubble exited the conical channel. Self-injected bunches are shown as colored dots with different colors and shapes (also for better readability of the graph). These figures are necessary to determine the optimal position of the self-injected bunch depending on the radius of the small base of the cone. For example, in the case of $r_{cone}$=2.16 $c/\omega_{pe}$, the position of the self-injected bunch is observed in the acceleration phase of the wakefield wave. The holding of the self-injected bunch is ensured by the compression of the wake bubble, which moves along the conical channel.

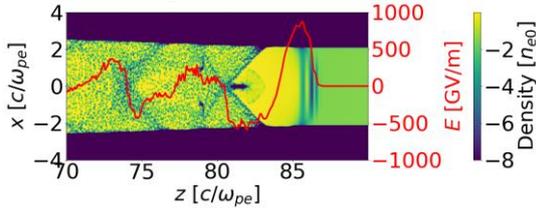

*Fig. 4. Density graph $n_e(z, x)$, acceleration field $E_z(x)$. Conical channel. $r_{cone}$=2.16 $c/\omega_{pe}$. t=238 fs.*

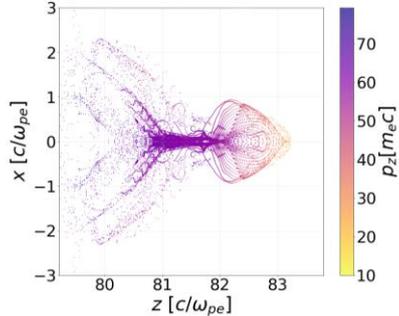

*Fig. 5. Longitudinal momentum $p_z(z, x)$. Conical channel. $r_{cone}$=2.16 $c/\omega_{pe}$. t=238 fs.*

Based on the analysis of Fig. 3, we will consider the radius of the conical channel to be optimal from the point of view of the balance of the values of the longitudinal momentum, the size of the bunch, the charge, and the accelerating field in the area of the bunch. Figure 6 shows the case of a cylinder. In Fig. 4, at the time of 238 fs, the mean longitudinal momentum of the self-injected bunch is $p_z$=57.6 $m_ec$.

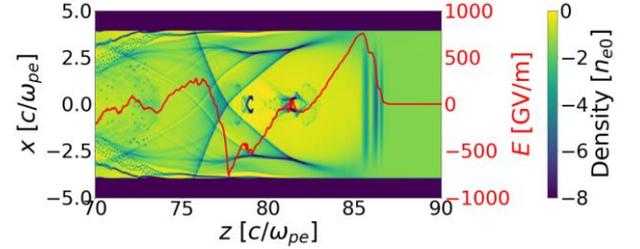

*Fig. 6. Density graph $n_e(z, x)$, acceleration field $E_z(x)$. Cylindrical channel. $r_{cyl}$=4.00 $c/\omega_{pe}$. t=238 fs.*

Fig. 7 shows the distribution of longitudinal momentum $p_z$. Fig. 6 and Fig. 7 show a self-injected bunch in a cylindrical channel, the average longitudinal momentum of which is equal to $p_z$=41.4 $m_ec$.

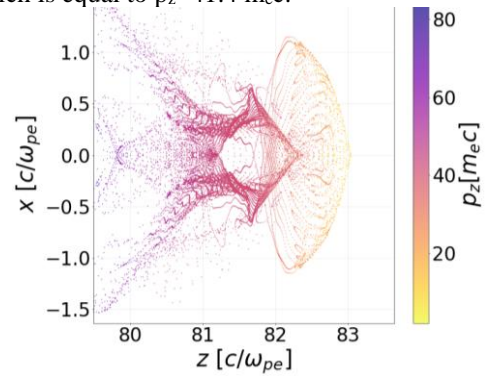

*Fig. 7. Longitudinal momentum $p_z(z, x)$. Cylindrical channel. $r_{cyl}$=4.00 $c/\omega_{pe}$. t=238 fs.*

Fig. 8 and Fig. 9 show the longitudinal momentum distributions for a conical channel (Fig. 8) compared to a cylindrical channel (Fig. 9). Approximately 1.6 times more particles are concentrated in the maximum region for a conical channel.

Thus, in case of conical channel ($r_{cone}$=2.16 $c/\omega_{pe}$) the average value of longitudinal momentum exceeds the average value in the case of a cylinder channel by 39%.

In the conical case, compared to the cylindrical case, the charge increases from 112.1 pC to 424.2 pC (3.78 times growth). The deviation of the longitudinal momentum $p_z$ varies from the mean from 8.2% at a radius of 1.26 $c/\omega_{pe}$, to 24.5% at a radius of 3.06 $c/\omega_{pe}$, and to 21.2% in the case of a cylinder (r=4.00 $c/\omega_{pe}$).

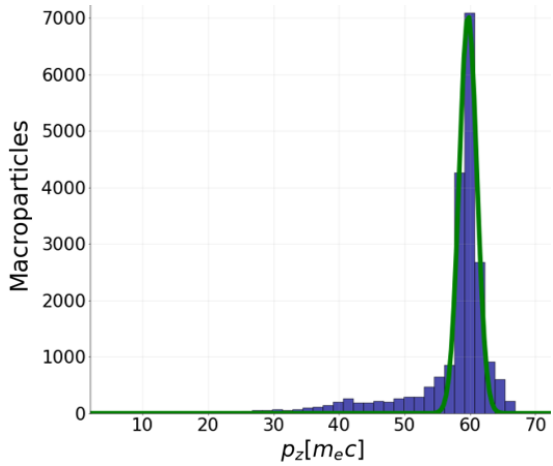

*Fig. 8. Longitudinal momentum $p_z$ distribution. Conical channel. $r_{cone}=2.16\ c/\omega_{pe}$. t=238 fs.*

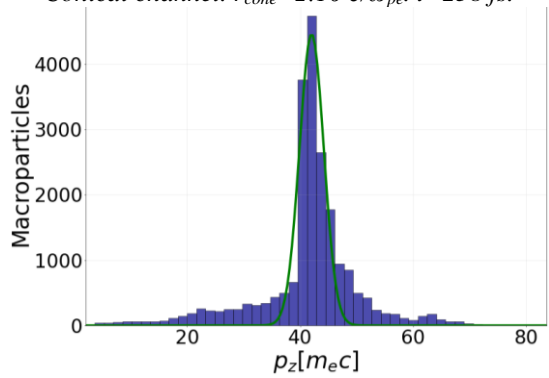

*Fig. 9. Longitudinal momentum $p_z(z, x)$. Cylindrical channel. $r_{cyl}=4.00\ c/\omega_{pe}$. t=238 fs.*

The average longitudinal momentum $p_z$ peaks at 58.8 $m_e c$ at r = 1.80 $c/\omega_{pe}$, before dropping to 42.0 $m_e c$ at r = 3.06 $c/\omega_{pe}$ and further to 41.2 $m_e c$ in the cylindrical case, highlighting the loss of acceleration efficiency beyond the optimal region (Fig. 2).

In Fig. 10 it can be seen that the use of a conical channel leads to compression of the laser pulse and growth of the field in the region of the beginning of the wake bubble at the end of the conical channel. The self-injected bunch leaves the phase of acceleration of the wake field, as a result of which a drop in the maximum field in the region of the bunch is observed with an increase in radius.

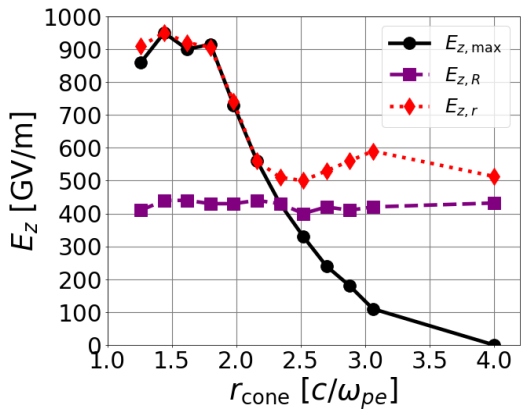

*Fig. 10. Dependence from small cone radius $r_{cone}$ (the cylinder is r = 4.0 $c/\omega_{pe}$) of the maximal acceleration field $E_{z,\ max}$ at the self-injected bunch region (black); maximal acceleration fields at the end of the wakefield bubble at the beginning ($E_{z,\ R}$ - purple) and at the end ($E_{z,\ r}$ - red) of the conical channel.*

## CONCLUSIONS

The authors completed a study of self-injected bunches and the accelerating field in the bunch region for a cylindrical channel and a set of conical channels with different radii of the small bases of the cone. The obtained dependencies themselves can be used for further studies. In comparison with the cylindrical channel, the conical channel ($r_{cone}$=2.16 $c/\omega_{pe}$) shows an increase in the longitudinal momentum of the bunch by 39%.

## ACKNOWLEDGMENTS

The study is supported by the National Research Foundation of Ukraine under the program "Excellent Science in Ukraine" (project # 2023.03/0182).

The authors are grateful to Leading Researcher V. A. Balakirev for productive scientific discussions.

**КІЛЬВАТЕРНЕ ПРИСКОРЕННЯ САМОІНЖЕКТОВАНОГО ЗГУСТКУ В ОДНОРІДНИХ КОНІЧНИХ ПЛАЗМОВИХ КАНАЛАХ**

*Д. С. Бондар, В. Ліманс, В. І. Маслов, І. М. Оніщенко*

Лазерне кільватерне прискорення - це широко досліджуваний метод прискорення згустків заряджених частинок, ключовою особливістю якого є самоінжекція. Однак, коли згусток прискорюється до швидкості, що перевищує швидкість драйвера, він зміщується фази максимального кільватерного поля прискорення. У цій роботі для вирішення цієї проблеми пропонується використовувати конічний канал, що звужується. Поступове звуження синхронізує згусток з фазою кільватерної хвилі, зменшуючи розмір бульбашки та утримуючи згусток у фазі прискорення. Порівняння з циліндричними каналами показує, що використання конічних каналів допомагає збільшити поздовжній імпульс згустка. Отримані корисні для подальших досліджень залежності довжини згустка, поля $E_z$, середнього поздовжнього імпульсу $p_z$ від радіуса каналу.